\UseRawInputEncoding

\documentclass[review,authoryear,12pt]{elsarticle}

\usepackage{etoolbox}
\makeatletter
\patchcmd{\ps@pprintTitle}
  {Elsevier}
  {journal (under review)}
  {}{}
\makeatother

\usepackage[english]{babel}
\usepackage{graphicx}
\usepackage{amsmath}
\usepackage{amssymb}
\usepackage{esvect}
\usepackage{mathtools}
\usepackage{algorithm} 
\usepackage{algpseudocode} 

\usepackage{hyperref}
\usepackage{caption}
\usepackage{float}
\usepackage{array}
\usepackage{bm}
\usepackage{optidef}
\usepackage{color}
\usepackage{amsthm}
\usepackage{comment}
\DeclareMathOperator{\sign}{sign}

\usepackage{natbib}
\usepackage{geometry}
\usepackage{fleqn}

\newtheorem{lem}{Lemma}
\newtheorem{prop}{Proposition}

\newtheorem{rem}{Remark}
\newtheorem{as}{Assumption}

\date{July 31, 2023}

\title{Robust data-driven learning and control of nonlinear systems. \\ A Sontag's formula approach\tnoteref{t1}}
\tnotetext[t1]{This work has been supported by the Project HOMPOT grant P20\_00597 under the framework PAIDI 2020 and by the National Program for Doctoral Formation (Minciencias-Colombia, 885-2020).}

\author[a,b]{Yeyson A. Becerra-Mora\corref{cor1}}
\ead{yeyson_becerra@cun.edu.co}
\author[a]{Jos\'e \'Angel Acosta}
\ead{jaar@us.org}

\cortext[cor1]{Corresponding author}
\affiliation[a]{organization={Dept. Ingenieria de Sistemas y Automatica, Universidad de Sevilla},
city={Sevilla},
postcode={41092},
country={Spain}}

\affiliation[b]{organization={Department of Electronic Engineering, CUN},
city={Bogota},
postcode={111711},
country={Colombia}}

\begin{document}

\begin{abstract}
An interlaced method to learn and control nonlinear system dynamics from a set of demonstrations is proposed, under a constrained optimization framework for the unsupervised learning process. The nonlinear system is modelled as a mixture of Gaussians and the Sontag's formula together with its associated Control Lyapunov Function is proposed for learning and control. Lyapunov stability and robustness in noisy data environments are guaranteed, as a result of the inclusion of control in the learning-optimization problem.
The performances are validated through a well-known dataset of demonstrations with handwriting complex trajectories, succeeding in all of them and outperforming previous methods under bounded disturbances, possibly coming from inaccuracies, imperfect demonstrations or noisy datasets. As a result, the proposed interlaced solution yields a good performance trade-off between reproductions and robustness.
The proposed method can be used to program nonlinear trajectories in robotic systems through human demonstrations. 
\end{abstract}

\maketitle

\section{Introduction}

Human-Robot collaboration has taken great relevance in recent years; nevertheless, programming a robot can be a difficult task for a non-expert user. Thus, an intuitive way to teach the robot a new specific task is by means of demonstrations. Taking into account that robotic paths are inherently nonlinear is necessary to propose new methods to estimate them. Learning from Demonstrations (LfD) \cite{r1} is a method that has been very well received recently to estimate (stable) dynamical systems (DS), hence, several works have used this approach.

Some works related to LfD have been characterized by the use of regression methods like Gaussian Process Regression (GPR), Locally Weighted Projection Regression (LWPR), and Gaussian Mixture Regression (GMR), among others. Another current is based on the optimization of policies as the Dynamic Movement Primitives (DMP) method. The approach presented in this work falls into the former.

Modeling nonlinear trajectories of DSs from user demonstrations is possible by employing Gaussian Mixture Model (GMM) and GMR \cite{r3}, with the combination of Hidden Markov Model (HMM) and GMR \cite{r4}, using GMM and HMM \cite{Vukovic}, or even employing GMM parameters into a Gaussian Process Regression (GPR) \cite{Wang}; nevertheless, neither of them ensures  stability. On the other hand, the Binary Merging method proposed in \cite{r5}, ensures local stability through a numerical process and employs GMM/GMR to build a trajectory estimation. Likewise, \cite{r6} use GMM to estimate a nonlinear point-to-point trajectory and define conditions to ensure Lyapunov stability; as well as \cite{r7} propose a constrained optimization problem to learn the unknown parameters of GMM that ensures asymptotic stability. In \cite{r8} a constrained optimization problem is employed to learn a Control Lyapunov Function (CLF) from demonstrations to ensure stability on an estimated DS. Following a similar approach, an improved estimate of the DS is presented in \cite{r9} which is based on transforming the data from demonstrations with a diffeomorphic candidate.
Diverse applications have been developed by using a DS estimate approach with regression methods and stability conditions such as: Obstacle avoidance, \cite{r10}; multiple attractors, \cite{r11}; capturing irregular objects in the air, \cite{r12}; rehabilitation therapy, \cite{r13}; visual serving, \cite{r14}; robotic handwriting, \cite{r15}; impedance control, \cite{r16}; virtual reality, \cite{r17}; among others.  

Other approaches for LfD have been addressed. Recurrent Neural Networks (RNN) have been used to learn a Lyapunov function from the user demonstrations like in \cite{r18}, where the learned function is used for stability. It is worth mentioning that this work and the one in \cite{r8} have resembled approaches, both of them divide the problem into two steps: Learning a function and using it to somehow stabilize the system. Nonlinear autoregressive polynomial models and a constrained least-square estimator are employed in \cite{r19} to learn a DS which is locally asymptotically stable. In \cite{Rego} a feed-forward neural network is employed to ensure stability and to estimate the region of attraction with the learned Lyapunov function. Finally, based on energy tank-based controllers, the work in \cite{r20} ensures the stability of a DS dissipating the energy. 

Our approach is inspired by two previous works, Stable Estimator of Dynamical systems (SEDS) \cite{r7} and Control Lyapunov Function-based Dynamic Movements (CLF-DM) \cite{r8}. The former learns optimal parameters from GMM to estimate a complex trajectory through GMR. In contrast, the latter learns an energy function and corrects the dataset using a virtual control in combination with a regression method, e.g. GMR. Moreover, SEDS learns the energy function and estimates the trajectory in one step, while CLF-DM do the same but in two different steps; achieving better accuracy in CLF-DM than in SEDS. In our study, we propose a hybrid learning algorithm that combines the advantages of SEDS and CLF-DM, additionally, improves the accuracy of estimates, and deals with noisy data. i.e. our method optimizes the GMM parameters and uses a control signal to enforce global exponential stability, and robustness. The method is validated in the LASA handwriting dataset \cite{r21}.

The paper is organized as follows. The problem statement and detailed contributions are in Section II; Section III summarizes GMM and GMR; Section IV is devoted to stability and robustness and Section V the learning. Section VI validates the performance with a dataset and Section VII conclusions and future research.

\noindent{\bf Notation.} $|\cdot|$ stands for vector norm, and $|\cdot|_{W}$ a weighted norm with matrix \mbox{$W\succ0$} where `$\succ$' stands for symmetric and positive definite. A function $\alpha(s) \in \mathcal{K}_{\infty}$ if $\alpha(0)=0$ and $\lim_{|x|\rightarrow \infty} \alpha(|s|)=~\infty$.

\section{Problem statement and contributions}

The underlying idea is to accurately estimate a `singular' (nonlinear) dynamics, which is represented through demonstrations, but where their model are completely unknown, because either are too complex to be obtained by first principles or the ultimate goal is to automate a complex task. Thus, only (noisy) input/output data are available through demonstrations.
First, let us be precise with the distinction between demonstrations and training sets used in the learning literature. Demonstrations reproduce trajectories of presumably stable (nonlinear) dynamics, while training sets do not necessarily. The proposed approach learns and controls jointly covering both. 
Let the trajectory to be learned the state vector $\bm x \in \mathbb{R}^{d}$, which encodes a set of $M$ demonstrations of $N$ data-points each.
Denote the pairs $\mathcal{D}:=\{x^{m,n}, \dot{x}^{m,n}\}_{m=1,n=1}^{M,N}$ the set of instances of an autonomous dynamical system given by
\begin{equation} \label{eq:sys}
\dot {x} = f(x) + u,
\end{equation}
where $f:\mathbb{R}^{d} \mapsto \mathbb{R}^{d}$ is a continuous and \emph{smooth} vector field, $u$ the control action of appropriate dimensions.
In order to facilitate notation, let us denote in `bold' each pair $\bm x^{m,n}:=\{x^{m,n}, \dot{x}^{m,n}\}$ and hence $\mathcal{D}:=\{\bm x^{m,n}\}_{m=1,n=1}^{M,N}$. The pair estimation mechanism is based on conditional probability using known data ${x}^{m,n}$ as prior probabilities
and posterior ones to estimate unknown $\dot{x}^{m,n}$. Thus, for example, for path planning control the pair encodes position and velocity, respectively. 
Let the learned controlled system be defined as
\begin{equation} \label{eq:sysu}
\hat {\dot {x}} = \hat f(x) + \hat u(x) + \eta(t,x),
\end{equation}
where $\hat f:\mathbb{R}^{d} \mapsto \mathbb{R}^{d}$ is a nonlinear estimate and $\hat u(x)$ the input estimate and $\eta$ a bounded additive disturbance accounting for inaccuracies in measurements, imperfect demonstrations due to bounded noise or even external disturbances coming from the original dynamics \eqref{eq:sys}. 
Without any loss of generality, we consider the target at $x^{*}=0$, and hence $f(0)=\hat f(0)=0$ as in \cite{r7}.
Notice that definition \eqref{eq:sys} accounts also for uncontrolled (open-loop) systems $u=0$, and hence stabilizable\footnote{The system \eqref{eq:sys} is stabilizable if for any $x(0)$ there exists $u(t)$ such that $\lim_{t\rightarrow \infty} x(t) = x^{*}$, $t\geq 0$.}
ones, but not necessarily open-loop stable. Therefore, the definition~(\ref{eq:sysu}) allows the estimated $\hat f(x)$ to represent open-loop unstable dynamics, unlike the related works. This is the prelude to the following and only assumption of the original dynamics \eqref{eq:sys}.

\begin{as} \label{as}
The system \eqref{eq:sys} is stabilizable at the target.
\end{as}

Once the problem is posed, we are in position to describe the contributions with respect to related works in these directions: {\bf C1} Stabilizability; {\bf C2} Optimization and {\bf C3} Robustness.

\noindent{\bf C1.} The closest and seminal related works which have served as inspiration for the present one are \cite{r7} and \cite{r8}. First, with respect to the class of systems, in there the class of systems and so its estimate are restricted to be stable, which is in fact a hard constraint imposed in the optimization problem. In here, Assumption \ref{as} does indeed relax that condition by requiring only stabilizability rather than stability of \eqref{eq:sys}. To give a simple example, consider the dynamics $\dot x = x + u$ as \eqref{eq:sys}, which is an open-loop unstable system and hence the approach \cite{r7} or \cite{r8} cannot be used. However, it is stabilizable and thus the approach proposed here can learn and estimate $\hat f(x)$ and $\hat u(x)$, ensuring stability and convergence to the target.

\noindent{\bf C2.} In \cite{r8}, the approach is split into two independent steps: 1) an estimate of a Lyapunov function, namely $V(x)$, in agreement with the demonstration set, and 2) an out-the-loop correction of such set of demonstrations through $u(x)$ to enforce the already estimated $V(x)$ to be a Lyapunov function for `mislearned' points.
To avoid any misunderstanding, let us clarify that although the latter might seem a remedy to impose stability, the fact is that in \cite{r8} the learning strategy priories estimation over execution, unlike here.
Essentially, they discard and replace the `mislearned' points of $\hat f(x)$ while keeping the solution $V(x)$ obtained in an decoupled optimization problem. Certainly, the main advantage is to solve two simplified and independent optimization problems for $V(x)$ and $\hat f(x)$, however, the disadvantage is that $V(x)$ might turn out not to be an accurate Lyapunov function for the original set of demonstrations and then the authors propose such suitable correction.
Unlike here, we propose to solve only one optimization problem in the loop for the demonstration set including $V(x)$, $\hat f(x)$ and $\hat u(x)$, which indeed is computationally lighter because it does not force the stable constraint, thus allowing to be used directly in real time for dynamics of the form \eqref{eq:sys}, i.e. without any low-level controller.

\noindent{\bf C3.} In this approach we consider undesirable disturbances all collected in $\eta$ of \eqref{eq:sysu}, which is a major difference with related works. The definition of the dynamics to be learned \eqref{eq:sys} considering the drift as the sum of $f(x)$ and $u(x)$ allow us to guarantee robustness. The disturbances can come from very different sources: either inaccuracies in measurements, imperfect demonstrations due to bounded noise and even external disturbances coming from the original dynamics. As a result, we prove that the inclusion of the appropriate $u(x)$ in the learning-optimization problem guarantees good performance trade-off between reproductions and robustness in noisy data environments.

\section{Regression via GMM and GMR}

The trajectory estimate \(\hat{f}\) from a set of demonstrations $M$ is defined as a mixture modeling with a finite set of Gaussian kernels, $K$. Mixture modeling builds a (coarse) representation of data density through a fixed number of mixture components. Methods like BIC \cite{Schwarz}, AIC \cite{Akaike}, DIC \cite{Spiegelhalter} can be used to find an optimal number of components. GMM is defined as an unsupervised learning or clustering algorithm which is based on Expectation Maximization (EM) to optimize the parameters of the Gaussian mixture \cite{r22}, likewise, EM initializes its parameters through K-means\cite{Bishop}. Although there are optimal regression methods like Gaussian Process \cite{Rasmussen}, they suffer the curse of dimensionality, i.e. the number of data points force to a linear growing pattern in the regressor.

Thus, GMM is able to construct the estimate $\hat f$ with the solely unknown parameters given by priors \(\pi_k\), means \(\mu_k\) and covariance matrices \(\Sigma_k\) of each Gaussian kernel $k$, where\footnote{Note that $\bm \theta_k$ is defined as a collection of vectors and matrices. Its dimension can be stated as $\dim(\bm \theta_k)=2 (K \times d^{2})$.} ${\bm \theta_k}:=\{\pi_k,\mu_{k},\Sigma_{k}\}$
 with $k=1,...,K$, and
\begin{equation}
    \mu_k = 
    \begin{bmatrix}
        \mu_{k}^{x}
        \\
        \mu_{k}^{\dot{x}}
    \end{bmatrix}\textrm{,} \quad
    \Sigma_k = 
    \begin{bmatrix}
        \Sigma_{k}^{x} & \Sigma_{k}^{x\dot{x}}
        \\
        \Sigma_{k}^{\dot{x}x} & \Sigma_{k}^{\dot{x}}
   \end{bmatrix}.
\end{equation}
Let $[{\bm x^{m,n}}]=[x^{m,n},\dot{x}^{m,n}]$ denote each data-point in the trajectory. Hence, in GMM the probability that a datapoint fits is defined by conditional probability density function $\mathcal{P}(\bm x^{m,n}|k)$ as a Gaussian normal distribution given by
\begin{equation*}
\mathcal{P}(\bm x^{m,n}|k) = \mathcal{N}(\bm x^{m,n}; \mu_{k}, \Sigma_{k}) =
{\left({(2\pi)}^{2d}{\rm det}({\Sigma}_k)\right)}^{-\frac{1}{2}}\exp{\left(-\frac{1}{2} |\bm x^{m,n}-\mu_k |_{(\Sigma_k)^{-1}}\right)},
\end{equation*}
and the mixture probability density function becomes 
\begin{equation*}
P({\bm x^{m,n}}; {\bm \theta_k}) = \sum_{k=1}^{K}\pi_k \mathcal{P}({\bm x^{m,n}} |k)
\
    \begin{dcases}
        k \in 1,\ldots,K,
        \\
        n \in 1,\ldots,N,
        \\
        m \in 1,\ldots,M.
    \end{dcases}
\end{equation*}
An estimated velocity is described through GMR \cite{r23} as
\begin{equation} \label{eq:fest}
    \hat{f}(x) = {\sum_{k=1}^{K} \gamma_{k}(x) (\mu_{k}^{\dot{x}}+\Sigma_{k}^{\dot{x}x}(\Sigma_k^{x})^{-1}({x}-\mu_{k}^{x}))},
\end{equation}
where we have defined the nonlinear weighting term \cite{Bishop} that measures the influence of the different Gaussians as
$$
\gamma_{k}(x):=\frac{\pi_k P({ x}|k)}{\sum_{i=1}^{K}\pi_i P({ x}|i)} \in \mathbb{R}_{[0,1]} .
$$

Even though $\hat{f}(x)$ can follow nonlinear trajectories, it is not possible to ensure asymptotic stability at the target with this method. However, Section IV presents the control strategy to deal with that. 
\begin{figure*}[t]
    \begin{center}
        \includegraphics[width=0.85\textwidth]{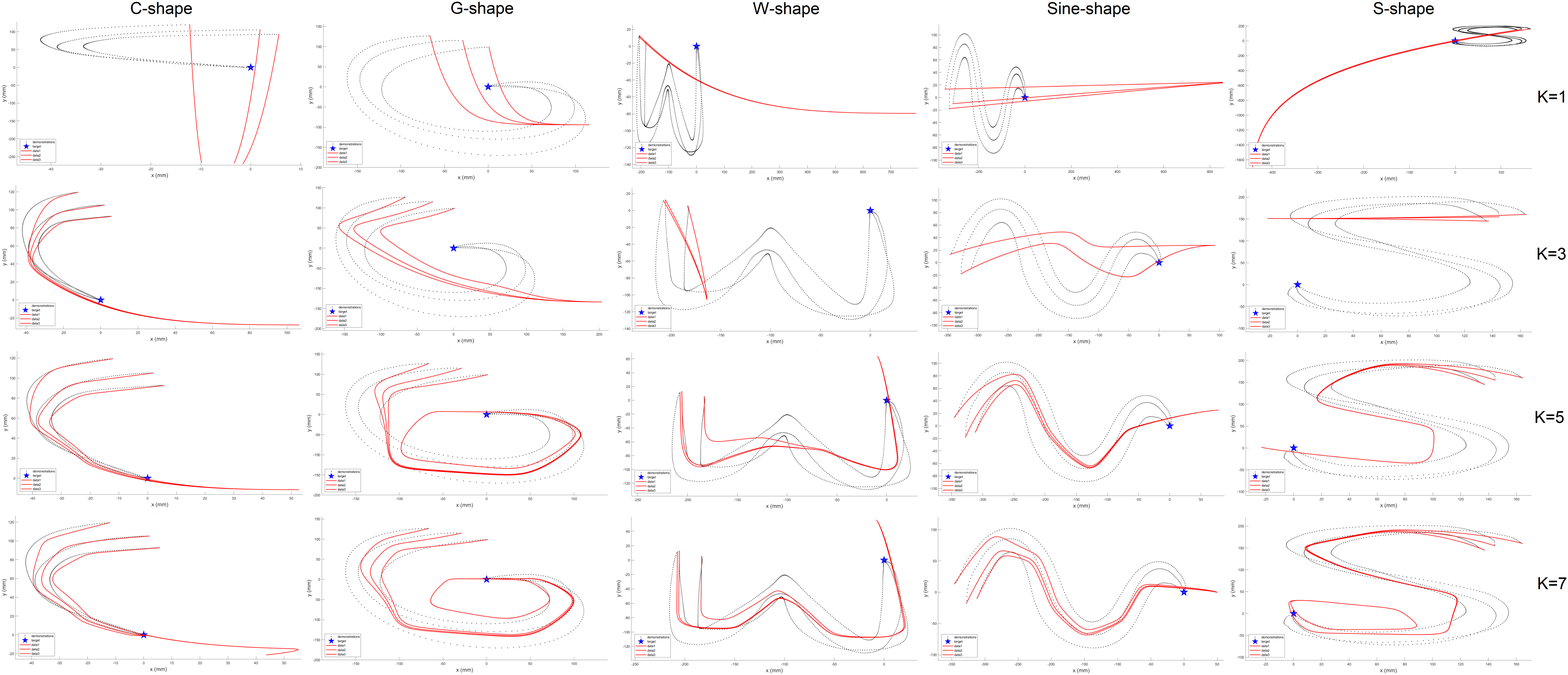}
        \caption{\small \sl Demonstrations (dotted black lines), estimates (red lines) and target (blue star)} 
        \label{fig:GMR}
    \end{center}
\end{figure*}
In Fig. \ref{fig:GMR} we show the estimate of GMR for different curve types C, G, W, Sine and S named by their shapes.
Even though, as $K$ increases---shown downwards in rows of Fig. \ref{fig:GMR}---, the estimates of the nonlinear trajectories are more similar to the demonstrations, simple shapes like C need a smaller $K$ than complex like W to find a better fit of the trajectory. In any case, note that no trajectory reaches the target, as expected. To enforce this ultimate goal some kind of correction is needed.

\section{Robust data-driven control with Sontag's formula}

For the control action we select the universal Sontag's formula which is the main difference with related works. Recall that the underlying idea is the parity between the existence of Lyapunov functions and the stabilisability with some special properties introduced in the work of Artstein \cite{Arstein} and subsequently in the work of Sontag \cite{Sontag}.
Artstein proved that there exists a stabiliser that makes the origin of \eqref{eq:sysu} asymptotically stable if and only if there exists a $\mathcal{C}^{1}$ Lyapunov function $V(x) > 0$, with $V(0) = 0$, satisfying the inequality
\begin{equation} \label{clf-opt}
\inf_{u\in \mathcal{U}} \ \{\nabla_x^\top V(x) (\hat{f}({x}) + \hat u(x))\} < 0 
\end{equation}
for at least a $u$ if $x \neq 0$ with $\mathcal{U} \subseteq \mathbb{R}^{l}$. Sontag named such Lyapunov function CLF. Artstein also proved that, although smooth elsewhere, such $u$ is not necessarily continuous at the origin, and provided this necessary and sufficient condition on $V$ to make $u$ continuous: for every $\epsilon>0$ there is a $\delta>0$ that whenever $|x| < \delta$, $ x \neq 0$, there is some $u$ with $|u|<\epsilon$ such that $\nabla_x V^\top (\hat{f} + \hat u) < 0$ holds, named is the small control property later on.
Defining $\bm a(x):=\nabla_x V(x)^{\top} \hat{f}({x})$, $\bm b(x):=\nabla_x V(x)^{\top}$ and ${\overrightarrow{\bm b}(x)}:=\bm b(x)^\top/|\bm b(x)|^{2}$, an analytical solution of the optimal control problem \eqref{clf-opt} reads

\begin{equation}\label{eq:uopt}
    {\hat u(x)=} 
    {\begin{dcases}
        - (\bm a(x) + \rho(| x|)) {\overrightarrow{\bm b}}(x), & \bm a(x) + \rho(|x|) > 0, \\
        0, & \bm a(x) + \rho(|x|) \leq 0,
    \end{dcases}} 
\end{equation}
where we dropped the arguments for compactness.
Notice that \eqref{eq:uopt} is a simplified version of the original Sontag's formula. Certainly, the Lyapunov function to be learned has a global minimum in $x^{*}=0$, which means that $\bm b(x^{*})=0$ and $\bm b(x) \neq 0$ for $x \neq x^{*}$, and hence ruling out the case $\bm b(x)=0$ in the original formula. More importantly, Sontag proposed as optimal\footnote{In \cite{Sontag} the control gain $\rho_{0}=1$, the adapted case for $\rho_{0}>0$ is in \cite{Acosta}.} $\rho(|x|) = \rho_0 \sqrt{\bm a(x)^{2}+|\bm b(x)|^{4}}$, $\rho_0>0$, but in general it could be any  positive one with $\rho(0)=0$. This is the main difference with \cite{r8} where minimum or no control effort was preferred forcing the $0$-case in \eqref{eq:uopt} and class $\mathcal{K}$ function  $\rho(|x|) = \rho_0(1-e^{-\kappa_0|x|})$, $\rho_{0},\kappa_{0}>0$, losing the robustness that we prove in this work.

\begin{figure*} 
    \centering     
    \includegraphics[width=0.85\textwidth]{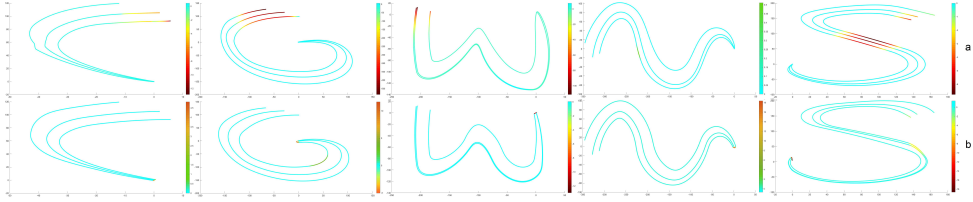}    
    \caption{\small Control Effect: a) our approach $\rho(|x|) = \rho_0 \sqrt{\bm a(x)^{2}+|\bm b(x)|^{4}}$ and b) $\rho(|x|) = \rho_0 (1-e^{\kappa_0 |x|})$ in \cite{r8}}
    \label{fig:control}
\end{figure*}
In any case, as in any Lyapunov control design for nonlinear systems the finding of a CLF is quite difficult. In this work, we fix the CLF as an asymmetric bi-quadratic function of the error as suggested in \cite{r8,r9}. It is readily seen that this is a limitation of the classes of systems that can be learned but it is kept in this work for comparison. Moreover, we also provide the stability vs robustness requirements unlike in there. Thus, let define $\sigma_{l}:\mathbb{R}^{d}\mapsto \mathbb{R}$ as $\sigma_{l}({ x}):={ x}^{\top} P_{l} ({ x - \mu}_{l})$, $\sigma_{l}(0)=0$, for $l=1,\dots,L$, and the CLF candidate  as
\begin{equation} \label{eq:V}
        V({ x}) = { x}^{\top} P_{0} { x} + \sum_{l=1}^{L} \sign_{+}(\sigma_{l}) \ \sigma_{l}({ x})^{2},
\end{equation}
with $\sign_{+}(\cdot)$ a coefficient defined as $0$ when its argument is negative and $1$ elsewhere.  Notice that \eqref{eq:V} is positive definite by definition upon $P_{0}>0$ irrespective of $P_{l}$. The  necessary requirements for stability are stated in the following lemma.

\begin{lem} \label{lem:V}
Consider $V:\mathbb{R}^{d}\mapsto \mathbb{R_{+}}$ defined in \eqref{eq:V} with $P_{l}\succ 0$, $l=1,\ldots, L$, then $V(0)=0$, $V({x})>0$ for all $x \neq 0$ and ${x}={0}$ is a global minimum.
\end{lem}
\begin{proof} The gradient and Hessian of \eqref{eq:V} become
\begin{align*}
\nabla_{x} V({x}) &= 2 P_{0}{ x} + 2 \sum_{l=1}^{L} \sign_{+}(\sigma_{l}) \sigma_{l}({ x}) P_{l} (2 { x - \mu}_{l}), \\
\nabla_{x}^{2} V({x}) &= 2 P_{0} 
+ 4 \sum_{l=1}^{L}  \sign_{+}(\sigma_{l}) \sigma_{l}({ x}) P_{l} \\ 
&+  2 \sum_{l=1}^{L} \sign_{+}(\sigma_{l}) P_{l} (2 { x - \mu}_{l}) (2 { x - \mu}_{l})^\top P_{l}.
\end{align*}
The necessary condition $\nabla_{x} V(0)=0$ comes from $\sigma_{l}(0)=0$. Noting that $\sign_{+}(\sigma_{l}) \sigma_{l}\geq 0$ and $\nabla_{x} V(x)^{\top} x>0$ by direct calculation, hence the Hessian is positive definite everywhere, which means that \eqref{eq:V} is globally convex.
\end{proof}
The function \eqref{eq:V} allows to model a broad set of complex functions, with the asymmetry provided by vectors \(\mu_l\) and the number of asymmetric functions $L$ defined by the user. 
In what follows in this section, we establish stability and robustness of the proposed approach, which optimizes while executing the feedback loop and is thus able to reject disturbances and hence diminishing the inevitable mismatch between the reproduction and the learned dynamics. 

The main result is stated in the following proposition.

\begin{prop} \label{pr:stab}
Consider the system dynamics \eqref{eq:sysu} with the state feedback \eqref{eq:uopt} under Assumption~\ref{as}. For any reproduction the following stability (i) and robustness (ii) results hold:
\begin{itemize}
\item [(i)] For $\eta=0$ the origin ${x} ={0}$ is a globally exponentially stable equilibrium for $\rho_{0}>0$ and $V(x)$ given by \eqref{eq:V} is a Lyapunov function;
\item [(ii)] Let $\eta, \dot \eta \in\mathcal{K}_{\infty}$, with $|\eta|^{2}< \kappa$ for some constant $\kappa>0$. Then closed-loop trajectories are globally uniformly ultimately bounded and converge to the residual set 
$$\Omega:=\left\{x\in\mathbb{R}^{d}: \rho(|x|) \leq {2 \kappa}/{\rho_{0}}\right\}.$$
\end{itemize}
\end{prop}
\begin{proof} Before proving the claims, the following facts are in order by construction:
\begin{itemize}
\item[F1.] By Lemma \ref{lem:V} the function \eqref{eq:V} satisfies $\nabla_{x} V({x})^{\top} x = \bm b(x) \ x > 0$ for $x \in \mathbb{R}^{d}/\{0\}$, and hence there exist functions $\underline{\alpha},\overline{\alpha}\in\mathcal{K}_{\infty}$ such that $\underline{\alpha}(|x|)\leq V(x) \leq \overline{\alpha}(|x|))$.
\item[F2.] By construction $\rho(0)=0 \Leftrightarrow x=0$ and $\lim_{|x|\rightarrow \infty} \rho(|x|)= \infty$, hence $\rho \in\mathcal{K}_{\infty}$.
\end{itemize}

\noindent{\bf Claim (i) [$\eta=0$].} The time derivative of \eqref{eq:V} along the trajectories of \eqref{eq:sysu} with \eqref{eq:uopt} becomes
\begin{align*}
\dot V({ x}) &= \nabla_{x} V({x})^\top 
\hat{\dot{x}} \\ 
&= \nabla_{x} V({ x})^\top \left( \hat f(x) - (\bm a(x) + \rho(|x|)) {\overrightarrow{\bm b}}(x)\right) \\
&= - \rho(|x|),
\end{align*}
and then F2 implies $\dot V(x) < 0$, for $x\in \mathbb{R}^{d}/\{0\}$. The result follows, noting that from F1 $V(x)$ is globally positive definite and radially unbounded and $V(x)=0 \Leftrightarrow x = 0$, and therefore there always exists a constant $c>0$ such that $\dot V(x) < - c V(x)$ concluding the proof of this claim.

\noindent{\bf Claim (ii) [$\eta\neq 0$].}
Likewise, an estimate of the time derivative of \eqref{eq:V} for \eqref{eq:sysu} with \eqref{eq:uopt} becomes
\begin{align*}
\dot V({ x}) &= \nabla_{x} V({x})^\top 
\hat{\dot{x}} \\ 
&= \nabla_{x} V({ x})^\top \left( \hat f(x) + \eta(t,x) - (\bm a(x) + \rho(|x|)) {\overrightarrow{\bm b}}(x)\right) \\
&\leq - \rho(|x|) + |\nabla_{x} V({x})^\top \eta(t,x)|  \\
&\leq - \rho_0 \sqrt{\bm a(x)^{2}+|\bm b(x)|^{4}} + \rho_{0}\frac{|\bm b(x)|^{2}}{2} + \frac{|\eta(t,x)|^{2}}{2\rho_{0}}  \\
&= - \frac{\rho_0}{4} \sqrt{(2\bm a(x))^{2}+|\bm b(x)|^{4}} + \frac{|\eta(t,x)|^{2}}{2\rho_{0}} \\
&\leq - \frac{1}{2} \left(\frac{\rho(|x|)}{2} - \frac{\kappa}{\rho_{0}} \right),
\end{align*}
where in the fourth line we have made use of Young's inequality and done some rearrangements in the fifth one. Recalling F2, $|\eta|^{2}<\kappa$ implies that $\dot V(x) < 0$ for $x\in \mathbb{R}^{d}/\Omega$. Therefore, F1 and F2 together with the set $\Omega$ being compact, we conclude global boundedness of trajectories. Finally, the ultimately claim comes from F1 ensuring compact level-sets of $V(x)$ and $\dot V(x) < 0$ for $x\in \mathbb{R}^{d}/\Omega$, and for any initial condition,  i.e. uniformly, concluding the proof.
\end{proof}

As it is readily seen, the stability margin of (i) naturally rejects bounded disturbances, and it could be interpreted as an input-output feedback problem with additive model uncertainty.

\begin{rem}
The class $\mathcal{K}$ function $\rho(|x|) = \rho_0 (1-e^{\kappa_0 |x|})$, $\rho_0,\kappa_0>0$, proposed in \cite{r8} does not provide any robustness to disturbances or inaccuracies. Moreover, in \cite{r8} the authors seek for minimal or no interaction of virtual control ($\rho$ above) to guarantee good reproductions. However, in this work we demonstrate that a good trade-off is achieved between robustness and good reproductions. To show the difference in Fig. \ref{fig:control} the control effect along the trajectories in all shapes is presented, note that demonstrations are free of noise. However, a trade-off is always needed and a discussion about this is carried out at the end of Section VI.
\end{rem}

\begin{figure*}
    \begin{center}         
        \includegraphics[width=0.85\textwidth]{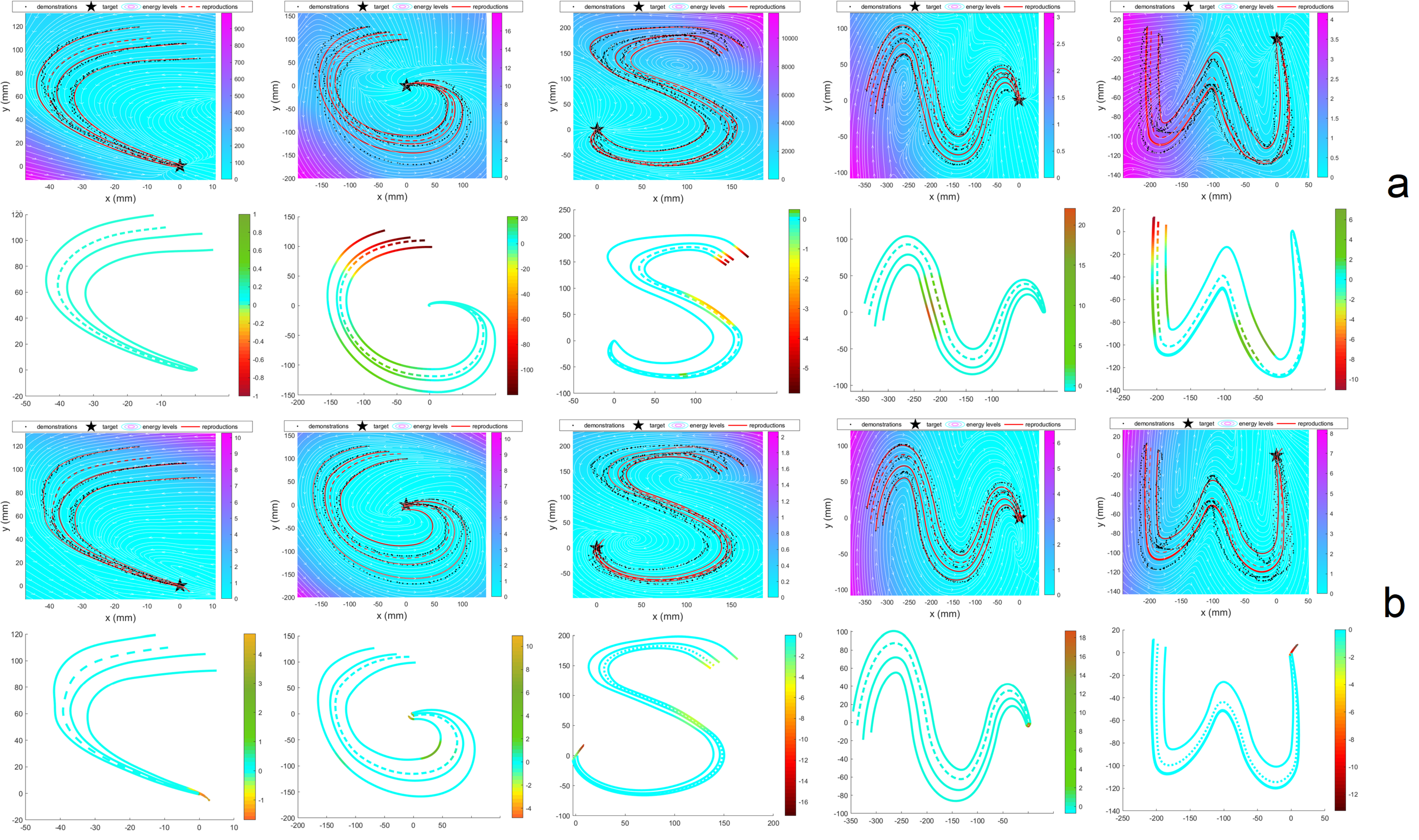}        
        \caption{\small Trajectories with a bounded disturbance: a) CLFSE (our Method) and b) CLF-DM Method. Demonstrations are depicted by dotted black lines and estimations by red lines. The Lyapuov function is depicted with color gradient} 
        \label{fig:noise}
    \end{center}  
\end{figure*}

\section{Learning by optimization algorithm}

In this section, we define a learning algorithm to determine the optimal parameters which is based on a constrained optimization problem. Recall that the parameters are ${\bm \theta_k}:=\{\pi_k,\mu_{k},\Sigma_{k}\}$  for GMR, and $P_l$ and \(\mu_l\) for the energy function. Let us define the whole set of parameters as $\bm \theta:=\{\bm \theta_k,P_l,\mu_l\}$.
The parameters for GMR are initialized with GMM which in turn uses EM, while the parameters to estimate the energy function, $P_l$ and \(\mu_l\) are initialized with identity matrices and null vectors, respectively.
We propose an eager learning in which the algorithm tries to optimize all parameters at the same time, unlike \cite{r8}. Even though this is an ambitious approach, we manage to consolidate 2 independent steps into just one, mainly because we removed the hardest stability constraint (20b) of \cite{r7}. However, the number of parameters to be optimized can increase considerably due to the number of Gaussian kernels $K$ and asymmetric quadratic functions $L$ utilized to estimate the nonlinear trajectory.
The objective function is defined as the mean square error (MSE) between the data and the estimates. The employed method to solve the constrained optimization problem is interior-point \cite{Wright}. Nevertheless, the solver cannot guarantee to find the optimal solution to this non-convex problem, but a local minima can be found through this method. 
Thus, consider a given data set ${\bm x}^{m,n}\in\mathcal D$, the constrained optimization problem becomes

\begin{mini}|s|
{\bm \theta}{J({\bm x}^{m,n}; \bm \theta) :=\frac{1}{2 M N} \sum_{m=1}^{M} \sum_{n=1}^{N} | \dot{x}^{m,n}-\hat{\dot x}^{m,n} |^{2}}
{\label{J}}{}
\addConstraint{\Sigma_{k=1}^{K} \pi^{k}}{=1;}{0 < \pi^{k}}{<1}
\addConstraint{\Sigma_{k}}{\succ 0,\quad}{k=1,\ldots,K}
\addConstraint{P_l + (P_l)^T}{\succ 0,\quad }{l=0,\ldots,L.}
\end{mini}

Algorithm \ref{alg} summarizes the optimization approach.
\begin{algorithm}
	\caption{CLF-based Stable Estimator} 
	\hspace*{\algorithmicindent} \textbf{Input:} $\bm {x}^{m,n}$, $\rho(|x|)$, $M,N,K$ and $L$
	\begin{algorithmic}[1]
	    \State Run GMM to initialize $\pi_{k_0}$, $\mu_{k_0}$ and $\Sigma_{k_0}$
	    \State Initialize $P_{l_0}$ and $\mu_{l_0}$
	    \While {$J>$ threshold and \eqref{J} are not satisfied}
	    \State Estimate $\hat{\dot{x}}^{m,n}$ from GMR $\bm \theta_{k}$ using \eqref{eq:fest}
	    \State Estimate $V(x^{m,n})$ from $P_L$ and $\mu_L$ using \eqref{eq:V} 
	    \State Compute  $\dot{V}(\bm x^{m,n})$ from \eqref{clf-opt}
	    \State Compute $u^{m,n}$ from \eqref{eq:uopt}
	    \If{$\dot{V}(\bm x^{m,n})$ $>$ 0}	     
		\ $(\hat{\dot{x}} + u^{m,n})$
	    \ElsIf {$\dot{V}(\bm x^{m,n})$ $\leq$ 0}	        
	    \ $\hat{\dot{x}}$
	    \EndIf
	    \State Minimize $|\dot{x}^{m,n} - \hat{\dot{x}}^{m,n}|^{2}$ of \eqref{J}
        \EndWhile 

	\end{algorithmic} 
	\hspace*{\algorithmicindent} \textbf{Output:} $\bm \theta^{*} =\{\pi_K^*, \mu_K^*, \Sigma_K^*, P_L^*, \mu_L^*\}$
    \label{alg}
\end{algorithm}

\section{Numerical Validation}

We validate the presented approach in a handwriting dataset \cite{r21}, and compare our results with SEDS \cite{r7} and CLF-DM \cite{r8} methods. The proposed experiments allow us to validate global exponential stability in the nonlinear system estimate (i.e. learning of complex trajectories) and to compare our results with previous methods; furthermore, we add a bounded disturbance to validate robustness in our approach. To carry on the experiments, five different shapes are chosen: C, G, W, Sine and S; which are based on its complexity to be estimated. Parameters $K$ and $L$ have to be carefully chosen to provide the most accurate estimate possible. 

\begin{figure}[htbp]
    \centering     
    \includegraphics[width=0.45\textwidth]{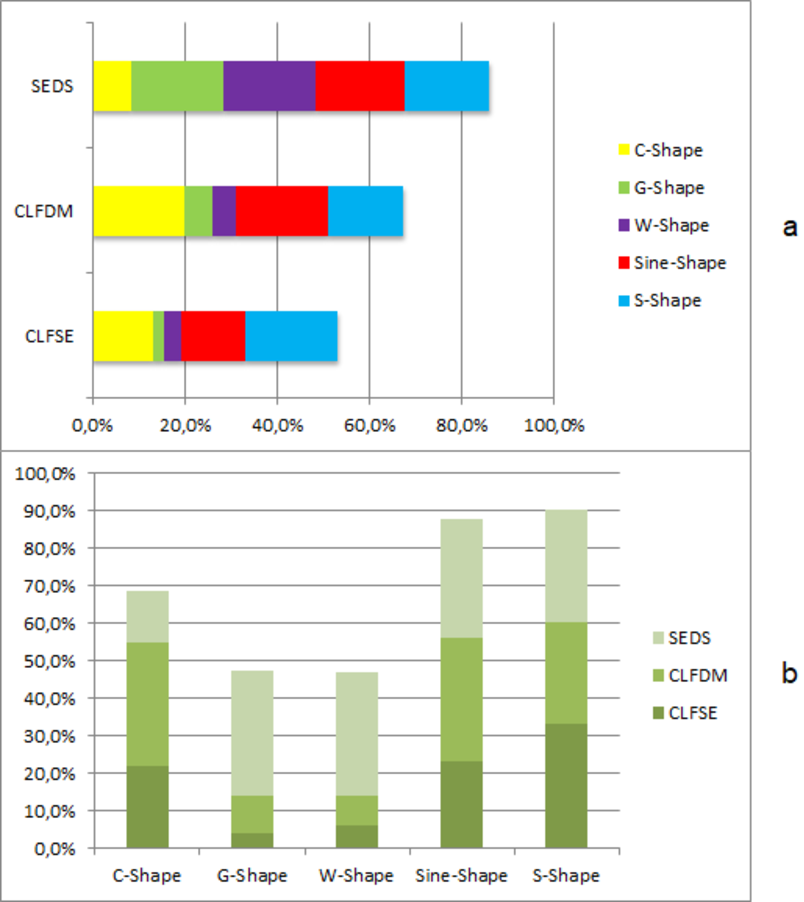}
    \caption{\small Swept Error Area among SEDS, CLFDM and CLFSE without disturbances}
    \label{fig:swept}
\end{figure}

As it was stated before, a regression method like GMR is employed to estimate complex trajectories. From Fig. \ref{fig:GMR}, an appropriate value for Gaussian functions can be observed; note that averaging the presented results in such a figure, a good estimate for the shapes is $K=5$, anyway simple shapes like C, can be estimated with a lower value (e.g. $K=3$). In section III, methods to determine an optimal value of $K$ are named; however, a simple way to determine $K$ is by testing how good the resulted estimation is. Nevertheless, GMR is not enough to ensure stability, and hence an asymmetric bi-quadratic function with the parameter $L$ has been used.

\begin{figure}[htbp]
    \centering  
    \includegraphics[width=0.45\textwidth]{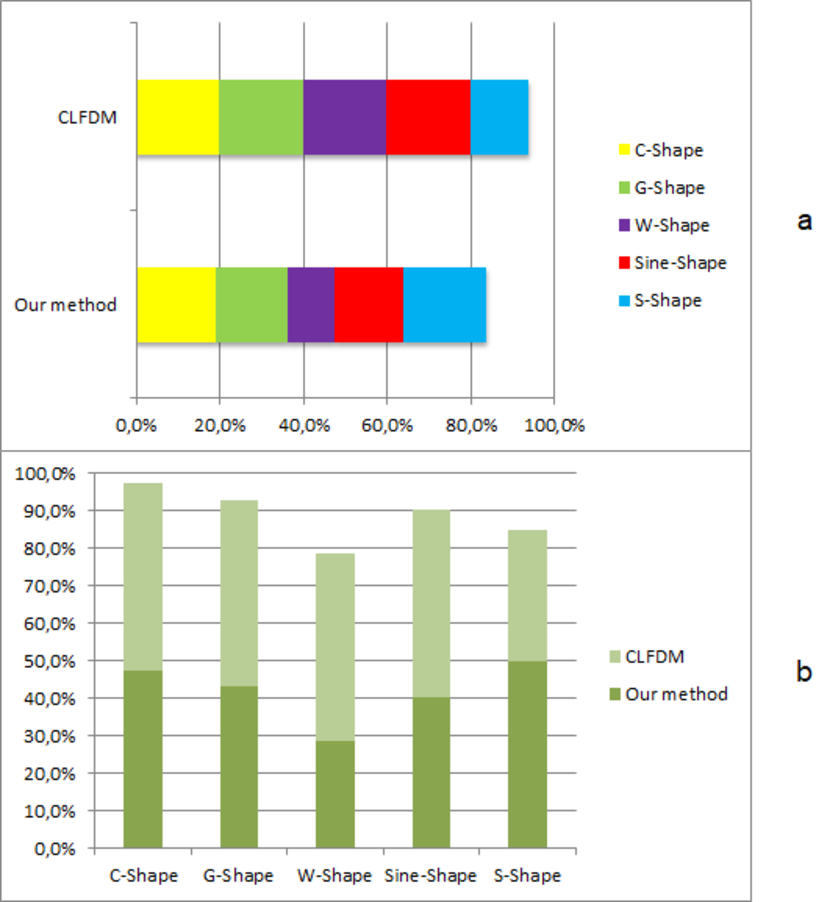}
    \caption{\small Swept Error Area between CLFDM and CLFSE with disturbances}
    \label{fig:swept-noise}
\end{figure}

On the one hand, the proposed method outperforms SEDS method \cite{r7} results due our estimates are more faithful to the demonstrations; in the other hand, it presents similar results with CLF-DM method \cite{r8} as it is showed in Fig. \ref{fig:control} and Fig. \ref{fig:noise}; nevertheless, our method keep presenting better results than CLF-DM. A way to quantify more precisely the difference between the demonstrations and estimates is through the swept error area, we validate that our method presents the minimum average error among all shapes and over all methods (see Fig. \ref{fig:swept}-a), and particularly outperforms to SEDS in more complex shapes like W or G (see Fig. \ref{fig:swept}-b). However, note that SEDS outperforms to CLF-DM and CLF-SE (our method) in C-shape, but it is because of the overparametrization, i.e. decreasing $K$ and $L$ in our method enhances the C-shape estimate.

Taking into account that CLF-DM method outperforms SEDS method in the previous experiment, we decide to compare our method and CLF-DM method in a second set of experiments. We add disturbances to the same shapes mentioned above. The disturbance levels employed for all shapes vary between 1\% and 5\% over the original data. Qualitatively, our method (see Fig. \ref{fig:noise}-a) outperforms CLF-DM method \citep{r8} (see Fig. \ref{fig:noise}-b) due all shapes converges to the target (look at the stream lines) and estimates are better. Note that our method is capable to employ a mayor control signal to overcome the disturbances, while CLF-DM method is not capable to increase the control signal effect to ensure ultimate boundedness. Furthermore, like the first experiment, we represent the results in a quantitative way through the swept error area. Once more, our method outperforms CLF-DM method results (see Fig. \ref{fig:swept-noise}). Additionally, we test our method with different initial conditions in order to prove the learning of the trajectory, see in Fig. \ref{fig:noise} the dotted red lines. 

\begin{figure}[htbp]
    \centering 
    \captionsetup{justification=centering}    
    \includegraphics[width=0.55\columnwidth]{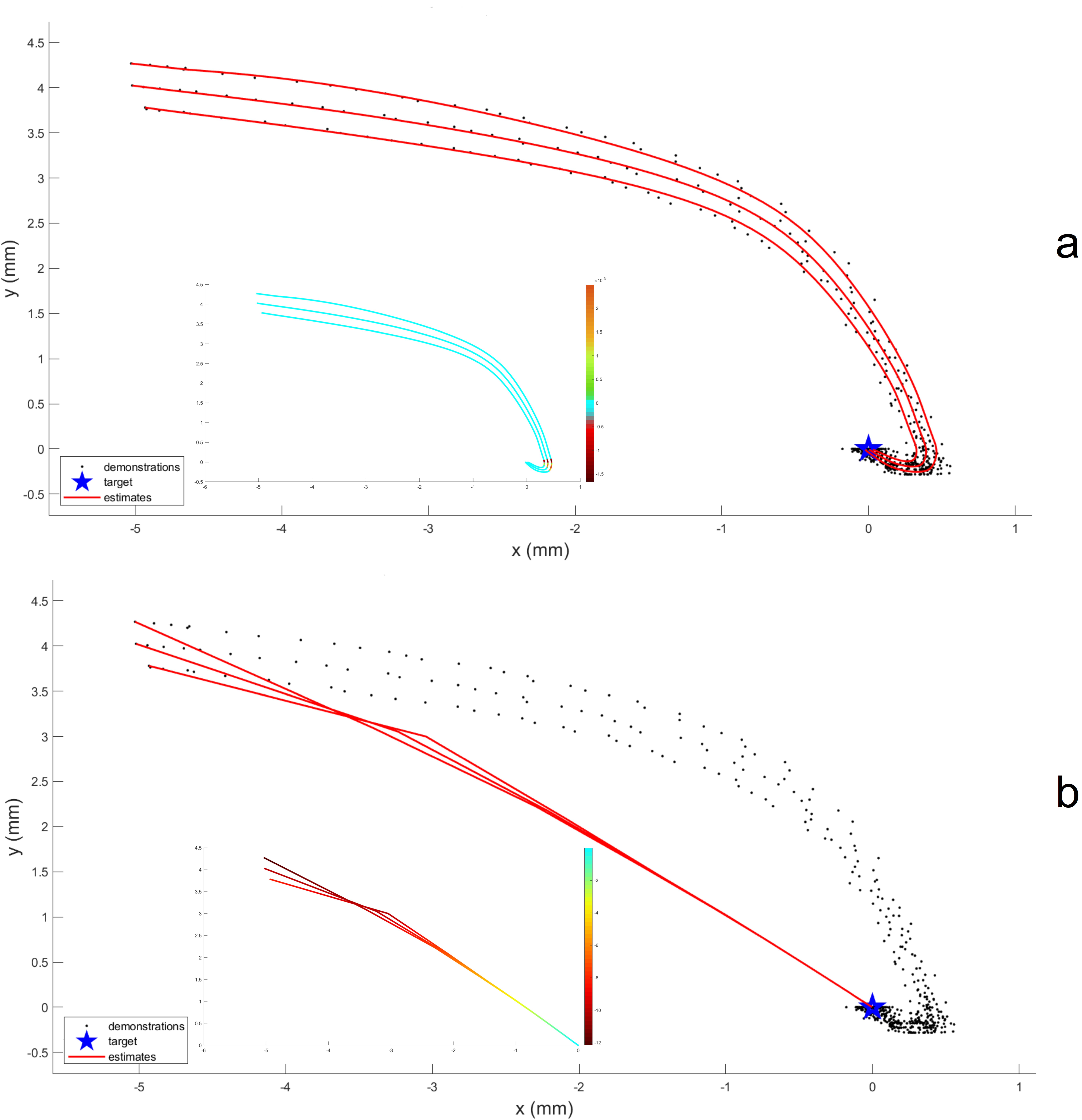}
    \caption{\small Comparison of $\rho_0$ parameter in a trajectory with disturbances.}
    \label{fig:rho}
\end{figure}

Finally, a comparison of $\rho_0$ parameter allows us to identify the relationship between the control signal effect and the regression. Thus, a better nonlinear trajectory estimate generates a least control signal effect, likewise a poor estimate takes to an increase in the control signal (see Fig. \ref{fig:rho}); therefore, as it was stated before, a trade-off between fidelity estimation and the control signal effect is needed.

\section{Conclusion}

In this work, we presented a method for unsupervised learning and control nonlinear systems ensuring (global) exponential stability and robustness. A novel optimization problem is provided to learn a Lyapunov-like function and GMR estimates complex trajectories from demonstration datasets. The method estimates controlled dynamical systems based on the mixture of Gaussians and Sontag's formula, and it learns through a constrained optimization problem. Compared with related methods, SEDS and CLF-DM, it solves the constrained optimization problem in one step, and guarantees stability and robustness to noisy datasets.
The validation of the approach is made through the available, and indeed very complete, handwriting dataset of different complex trajectories. The method success in all of them even with disturbances.
The proposed method outperforms former SEDS and CLF-DM, thus allowing its implementation in real time applications.
Current work is underway to address more complex trajectories that are difficult to learn like the W-shape, as well as to test our method in physical systems.

\section*{Acknowledgements}
The authors acknowledge support from the Project HOMPOT grant P20\_00597 under the framework PAIDI 2020  and by the National Program for Doctoral Formation (Minciencias-Colombia, 885-2020).

\bibliographystyle{elsarticle-harv}
\bibliography{ref}

\end{document}